\documentclass[twocolumn,showpacs,prl,amsmath]{revtex4}
\usepackage[dvips]{graphicx}
\usepackage{psfig,epsfig,pstricks}
\usepackage{comment}
\usepackage{amsmath,amsfonts,amssymb,amsthm}
\usepackage{xspace}
\usepackage[colorlinks=true,linkcolor=blue,citecolor=blue,filecolor=blue,urlcolor=blue]{hyperref}

\newcommand{\ie}{{\it i.e.}\xspace}

\newcommand{\Ito}{It\^{o}}
\newcommand{\gest}{g_{\text{est}}}
\newcommand{\eq}{\hspace{-.15cm}=\hspace{-.15cm}}
\newcommand{\ave}[1]{\left\langle#1 \right\rangle}

\newcommand{\elabel}[1]{\label{eq:#1}}
\newcommand{\eref}[1]{(Eq.~\ref{eq:#1})}

\newcommand{\Eref}[1]{Equation~(\ref{eq:#1})}

\newcommand{\flabel}[1]{\label{fig:#1}}
\newcommand{\fref}[1]{Fig.~\ref{fig:#1}}
\newcommand{\Fref}[1]{Figure ~\ref{fig:#1}}

\begin{document}

\title{Ergodicity breaking in geometric Brownian motion}
 
\author{\vspace{-.3cm}O. Peters}
\email{o.peters@lml.org.uk}
\affiliation{
London Mathematical Laboratory, 14 Buckingham Street, WC2N 6DF London, UK
}
\author{\vspace{-.3cm}W. Klein}
\affiliation{Physics Department and Center for Computational Science, Boston University, Boston, MA 02215}
\date{\today}
\begin{abstract} 
Geometric Brownian motion (GBM) is a model for systems as varied as
financial instruments and populations. The statistical properties of
GBM are complicated by non-ergodicity, which can lead to ensemble
averages exhibiting exponential growth while any individual trajectory
collapses according to its time-average. A common tactic for bringing
time averages closer to ensemble averages is diversification. In this
letter we study the effects of diversification using the concept of
ergodicity breaking.
\end{abstract}

\pacs{02.50.Ey,05.10.Gg,05.20.Gg,05.40.Jc
}
\maketitle 

{\it Geometric Brownian motion} (GBM) is a useful model for systems in
which the temporal evolution is strongly affected by relative
fluctuations, such as stock prices and populations. Fluctuations have
a net-negative effect on growth in such systems, due to the
multiplicative nature of the noise. One strategy commonly employed to
reduce these effects is diversification.  In the language of
statistical physics diversification involves a {\it partial ensemble
  average} (PEA) over a finite number, $N$, of trajectories generated
by GBM. This raises important questions such as: How do the PEAs
compare to the ensemble average ($N \to \infty$)?  How and when do
significant differences arise? In this letter we analyze PEAs of GBM
both analytically and numerically.

In GBM it is possible for the ensemble average to grow exponentially,
while any individual trajectory decays exponentially on sufficiently
long time scales \cite{Peters2011a}. Multiplicative growth is
manifestly non-ergodic. But precisely the opposite is often assumed in
economics, for instance in \cite{ChernoffMoses1959}, p.98: ``If a
gamble is `favorable' from the point of view of the expectation value
[ensemble average] and you have the choice of repeating it many times
[time average], then it is wise to do so. For eventually, your amount
of money [is] bound to increase.'' Some of the consequences of this
unwarranted assumption of ergodicity were pointed out in
\cite{Peters2011a}, here we treat the general case of PEAs for
arbitrary averaging time and sample size.

Geometric Brownian motion is defined by
\begin{equation}
dx=x(\mu dt + \sigma dW),
\elabel{GBM}
\end{equation}
where $\mu$ is a drift term, $\sigma$ is a noise amplitude, and
$W(t)=\int_0^t dW$ is a Wiener process. Without the noise, \ie
$\sigma=0$, the model is simply exponential growth at rate $\mu$. With
$\sigma\neq 0$ it can be interpreted as exponential growth with a
fluctuating growth rate.

To solve \eref{GBM}, one computes the increment $d\ln(x)$ of the
logarithm of $x$, integrates and exponentiates. The interesting step
is computing the increment because this requires stochastic
calculus. In writing \eref{GBM} we had in mind an interpretation of
the equation in the \Ito ~convention \cite{LauLubensky2007}. With this
convention, it is well known
that $d\ln(x)=(\mu-\frac{\sigma^2}{2})dt + \sigma dW$, which by
exponentiation implies the solution
\begin{equation}
x(t)=x(0) \exp\left(\left( \mu-\frac{\sigma^2}{2}\right)t+\sigma W(t)\right).
\elabel{solution}
\end{equation}
For simplicity, we will assume the initial condition
$x(0)=1$. \Fref{contours} illustrates the nature of this process.
\begin{figure}
\includegraphics*[width=\columnwidth,angle=0]{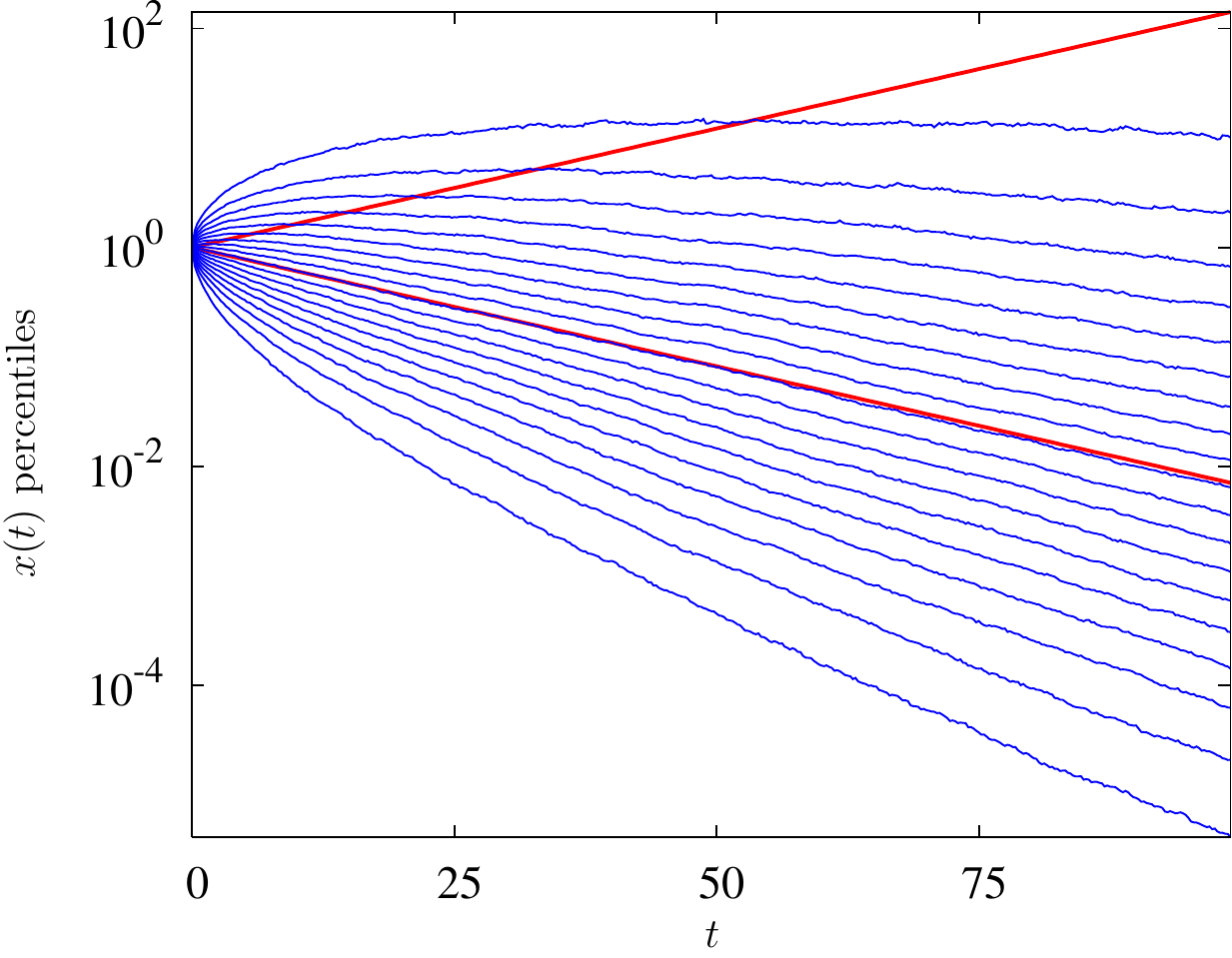}
\caption{Percentiles from 95 to 5 (top to bottom) in steps of 5, based
  on 10,000 realizations of $x(t)$. The parameters (used for all
  illustrations here) are $\mu=0.05$ and $\sigma^2=0.2$. The red
  straight lines show the ensemble average (upward sloping) and an
  exponential decreasing with the time-average growth rate. The
  ensemble average is essentially meaningless for a single
  realization, whereas the time-average growth rate accurately
  describes the typical behavior.}  \flabel{contours}
\end{figure}

The process $x(t)$ is not stationary.  This implies that where
averages can be defined, there is no guarantee for ergodicity, \ie the
equality of ensemble and time averages \cite{Gray2009}. The time
average of the process itself is
either 0 (if $\mu-\sigma^2/2 < 0$), or diverges positively (if
$\mu-\sigma^2/2 > 0$), whereas the ensemble average is an exponential
function of time. To capture the non-ergodicity of the process in
well-defined averages of an observable, we define the following
estimator for the exponential growth rate
\begin{equation}
g_{\text{est}}(t,N):=\frac{1}{t}\ln\left(\ave{x_i(t)}_N\right),
\elabel{estimator}
\end{equation}
where we call $\ave{\cdot}_N:=\frac{1}{N}\sum_i^N \cdot$ a PEA. The
estimator looks at the growth rate of a PEA (it is not a PEA of the
growth rate), \ie the logarithm is taken outside the average. This is
crucial to leave the non-ergodic properties of the process, $x(t)$,
intact. The time-average growth rate, denoted $\bar{g}$, is found by
letting time remove the stochasticity in the process. Mathematically,
this is the limit
\begin{equation}
\bar{g}:=\lim_{t\to \infty}g_{\text{est}}(t,N)=\mu-\frac{\sigma^2}{2}.
\elabel{time}
\end{equation}
The ensemble-average growth rate, denoted $\ave{g}$, is found by
letting an increasing ensemble size remove the
stochasticity. Mathematically, this is the limit
\begin{equation}
\ave{g}:=\lim_{N\to \infty}g_{\text{est}}(t,N)=\mu.
\elabel{ensemble}
\end{equation}
The non-ergodicity of the process is manifested in the
non-commutativity of the limits $\lim_{t\to\infty}$ and
$\lim_{N\to\infty}$.

Both ensemble and time averages are mathematical
objects and therefore separated from physical reality by the divide
that separates logic from matter. Nonetheless, both averages carry
practically meaningful messages. To identify the regimes where they
``apply'', that is, where they reflect typical behavior, it is
important to understand more about the general case where both the
observation time, $t$, and the ensemble size, $N$, are finite and
arbitrary.

In \cite{Peters2011a}, the time-average growth rate, \eref{time}, was
computed for a single system, $N=1$, by letting $t$ diverge.  This
case is related to the so-called Kelly criterion, a concept from the
gambling literature \cite{Kelly1956}, discussed in
\cite{Peters2011a,Peters2011b}. But the case of arbitrary $N$ was not
treated. The ensemble average was computed for arbitrary $t$. But the
limit $N\to \infty$ was not taken explicitly, relying on the fact that
in this limit the PEA, $\ave{x_i(t)}_N$, is the expectation value,
$\ave{x(t)}$. Below we show that \eref{time} holds for arbitrary
finite $N$ and characterize the process of the convergence of
\eref{ensemble} for arbitrary finite $t$ as $N\to\infty$.

We begin by showing that for a single instance, $N=1$, the
distribution of $\gest(t,N\eq1)$ approaches a delta function centered
on $\mu-\sigma^2/2$ in the limit $t\to \infty$.

Substituting \eref{solution} in \eref{estimator},
$\gest(t,N\eq1)=\mu-\frac{\sigma^2}{2}+\frac{1}{t}\sigma W(t)$. We
know that the distribution of $W(t)$ is Gaussian with mean 0 and
standard deviation $t^{1/2}$, which we write as
$P(W(t))=\mathcal{N}(W(t);0,t^{1/2})$. To compute the distribution of
$\gest(t,N\eq1)$, we use the transformation law of probabilities,
$P(g)=P(W)\left|\frac{dg}{dW}\right|^{-1}$. With
$dg/dW=\frac{\sigma}{t}$ and solving \eref{solution} for $W(g)$, this
yields
\begin{align}
P(\gest(t,N\eq1))
&=\mathcal{N}\left(\gest;\mu-\frac{\sigma^2}{2},\sqrt{\frac{\sigma^2}{t}}\right)\elabel{Gaussian}
\end{align}
The limiting behavior of this distribution for $t\to\infty$ is the
Dirac delta function
\begin{equation}
\lim_{t\to\infty}P(\gest(t,N\eq1))=\delta\left(\gest-\left(\mu-\frac{\sigma^2}{2}\right)\right)
\end{equation}
In other words as $t \to \infty$, the observed growth rate
will differ from $\mu-\frac{\sigma^2}{2}$ with probability zero.

Next, we consider $N$ instances of \eref{GBM}. At each
moment in time, the $N$ instances are averaged, as illustrated in
\fref{trajectories}, and we are interested in the long-time behavior,
$t \to \infty$, of the object $\ave{x(t)}_N$.

\begin{figure}
\includegraphics*[width=\columnwidth,angle=0]{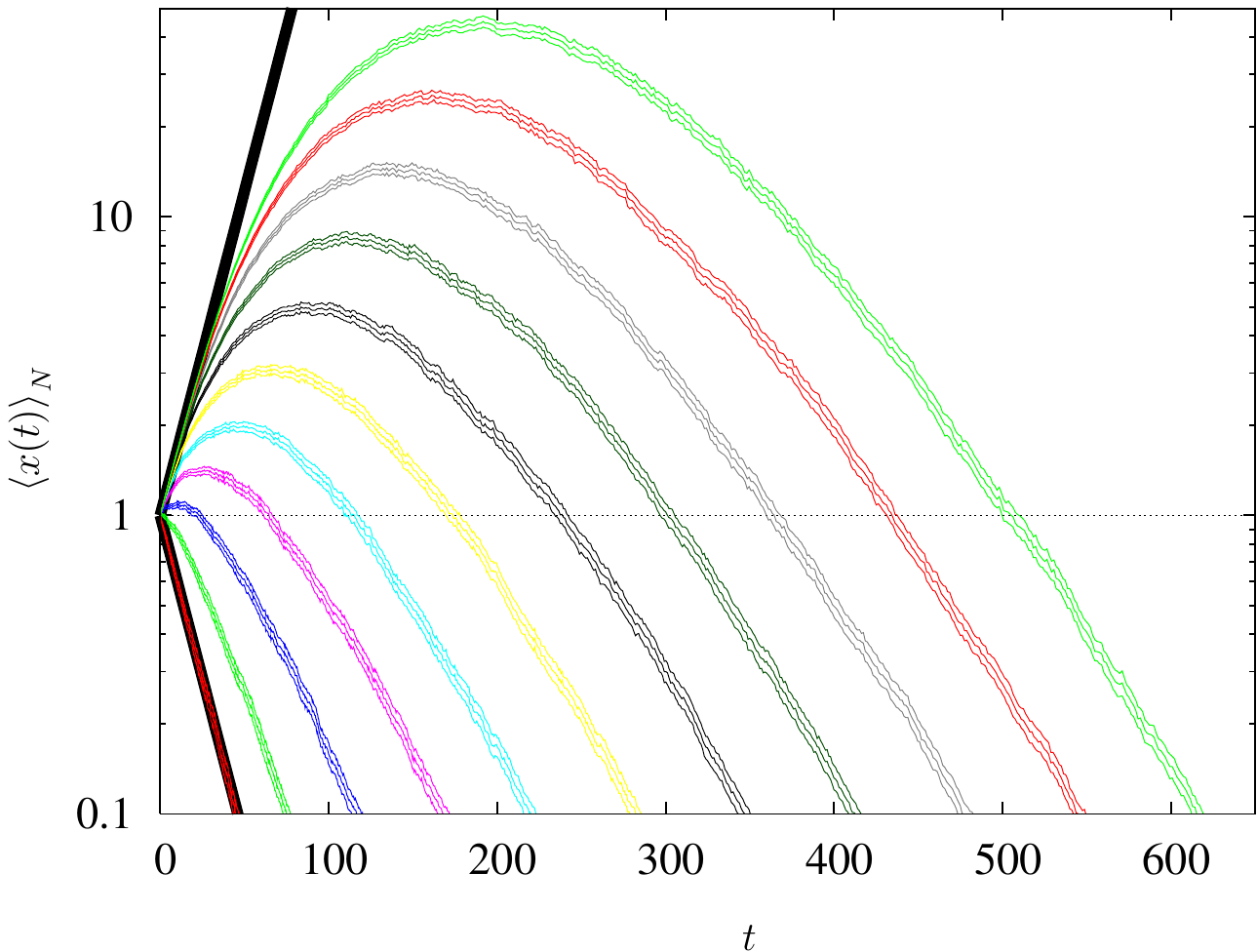}
\caption{Typical trajectories $\ave{x(t)}_N$ for different ensemble
  sizes $N$, from top to bottom $N=1024, 512...1$. For each ensemble
  size, 10,000 trajectories were generated. At each time the median,
  the 49th and 51st percentiles are shown. The solid straight red
  lines illustrate the ensemble-average and time-average growth
  rates. The single system (bottom) is dominated by time-average behavior;
  for increasing $N$ the trajectories stay close to the ensemble
  average.}  \flabel{trajectories}
\end{figure}

Again, substituting \eref{solution} in \eref{estimator},
\begin{align}
\gest(t,N)&=\frac{1}{t}\ln\left(\ave{
\exp\left(\left( \mu-\frac{\sigma^2}{2}\right)t+\sigma W_i(t)\right)}_N\right)\\
&=\mu -\frac{\sigma^2}{2}+\frac{1}{t}
\ln\left(\ave{
\exp\left(\sigma W_i(t)\right)}_N\right)
\elabel{correction}
\end{align}

The difficulty with this equation is the logarithm of an average of
exponentials. Unlike in the case of the single system, the logarithm
does not simply undo the exponential, and the non-trivial behavior of
typical trajectories of PEAs is a direct result.

To show that \eref{time} holds for arbitrary $N$, we will proceed in
two steps by showing that as $t\to\infty$ 
\begin{itemize}
\item 
the probabilty of finding $\gest(t,N)>\mu-\frac{\sigma^2}{2}$
approaches zero (upper bound)
\item the probabilty of finding $\gest(t,N)<\mu-\frac{\sigma^2}{2}$
  approaches zero (lower bound).
\end{itemize}

\underline{Upper bound:}\\ \Eref{correction} is a growth rate estimate
of an average. But the average cannot be larger than the largest
individual term. This establishes an inequality, namely an upper bound
on $\gest(t,N)$.
\begin{equation}
\gest(t,N)\leq\mu-\frac{\sigma^2}{2}+\frac{1}{t}\sigma \max_i^N W_i(t).
\end{equation}
A value $\gest(t,N)>\mu-\frac{\sigma^2}{2}+\epsilon$ is thus only
possible if
\begin{equation}
\max_i^N W_i(t) > \frac{\epsilon t}{\sigma}.
\end{equation}
The probability of such an extremum is \cite{Gumbel1958}
\begin{align}
P\left(\max_i^N W_i(t) > \frac{\epsilon t}{\sigma}\right)=1-\left(\int_{-\infty}^{\frac{\epsilon t}{\sigma}}\mathcal{N}(z;0,\sqrt{t})dz\right)^N.
\end{align}
Two interesting properties can be observed. First, for finite $N$,
and $\epsilon>0$,
\begin{align}
\lim_{t\to\infty}P\left(\max_i^N W_i(t) > \frac{\epsilon t}{\sigma}\right)=0,
\end{align}
the desired result. This is because the width of the distribution
$\mathcal{N}(x;0,\sqrt{t})$ increases as $\sqrt{t}$, whereas the upper
limit of the integral increases as $t$, outpacing the divergence of
the width. In the limit $t\to\infty$, the entire distribution
is integrated, which yields 1 due to normalization.  Second, for
finite $t$, the limit
\begin{align}
\lim_{N\to\infty}P\left(\max_i^N W_i(t) > \frac{\epsilon t}{\sigma}\right)=1.
\end{align}
This must be so because the term that is being raised to the
$N^{\text{th}}$ power is less than 1 for finite $t$ and therefore
vanishes exponentially with $N$. In other words, in the limit of
diverging ensemble size, the method fails to give an upper bound on the
estimated average growth rate. This is so because
$\max_i^N W_i(t)$ diverges in this limit.

\underline{Lower bound:}\\ The lower bound on $\gest(t,N)$ is obtained
in the same way as the upper bound, by switching the inequality and
considering the minimum of the $N$ instances at time $t$.

We have shown that as $t \to \infty$ the probability for observing a
growth rate $\gest(t,N)\neq \mu-\frac{\sigma^2}{2}$ approaches zero
for any finite $N$. 

In proving \eref{time}, we have used extreme values, and they will be
the key to understanding our problem: the exponential introduces a
weighting that leads to a finite contribution to the average from
extreme values whose relative frequencies vanish in the limit
$N\to\infty$. Considering the discrete-time, discrete-space random
walk, it is clear that the absolute maximum -- not the typical maximum
but the largest possible value -- scales in a light-cone fashion as
$t$ and not as $t^{1/2}$. This is reflected in two regimes of actual
physical diffusion, a short-time regime where extrema among the
positions of diffusing particles scale as $t$ and a long-time regime
where they scale as $t^{1/2}$, beautifully illustrated in
\cite{Richardson1921} and, using the discrete multiplicative binomial
process, in \cite{Redner1990}. For the Wiener process the largest
possible values are infinite for any $t>0$. This is a well-known
limitation of the model. As pointed out in \cite{MorseFeshbach1953},
the canonical solution to the diffusion equation violates special
relativity because it allows diffusing matter to exceed the speed of
light. This is visible in the small-$t$ behavior: while for large $t$
positions at a distance $\sqrt{t}$ correspond to slow motion
representing the physics well, at short times the speed of such motion
diverges and becomes unphysical.

In \cite{Redner1990} it was argued that in order to observe
ensemble-average behavior ($N\to \infty$) for a time $\tau$ in a PEA,
$N \sim \exp(\tau)$ multiplicative systems are required. This scaling
follows from the exponential decrease with $\tau$ of the probability
of $\tau$ consecutive up-moves in a random walk. The multiplicative
nature of the process enhances large outliers and leads to the extreme
values dominating the (linear) average behavior. After a time $\tau
\sim \ln(N)$ the absolute extremes become a-typical for the ensemble
size, leading to a deviation from
ensemble-average behavior.  The result in
\cite{Redner1990} is derived in the large $N$ limit. Here we
reconsider this problem, phrasing it in terms of the stability of
PEAs, and obtain a somewhat different conclusion. In particular we
find that the PEA deviates from the ensemble average at an earlier
time and is linearly unstable, \ie unstable with respect to
arbitrarily small perturbations coming from the noise.

We begin by defining the deviation $\epsilon_N(t)$ of
the PEA from the ensemble average by
\begin{equation}
\ave{x(t)}_N=\exp(\mu t)+\epsilon_N(t).  \elabel{epsilon} 
\end{equation}
Initially, trajectories will approximate those of the ensemble average 
so that we can approximate the deviation as
\begin{equation}
\epsilon_N(t)=\exp\left(\mu t + \sigma \ave{\int_0^{t_-}dW_i}_N\right) -
\exp(\mu t).
\elabel{lin_approx}
\end{equation}
Replacing $\ave{\int_0^{t_-}dW_i}_N$ by
$\sqrt{\ave{\ave{W_i(t)}_N^2}}=\sqrt{\frac{t}{N}}$ we obtain an
expression for the scaling behavior of the deviation
\begin{equation}
\epsilon_N(t) \sim \sigma \exp(\mu t) \sqrt{\frac{t}{N}}.
\elabel{epsilon_scaling}
\end{equation}
It can be shown that $\epsilon_N(t)$ is the PEA of the solution to
\begin{equation} 
d\epsilon_{N=1}(t) = dt \mu \epsilon_{N=1}(t) + \sigma \exp(\mu t)dW
\elabel{escale-eq}
\end{equation}
with the initial condition $\epsilon_N(t) = 0$ at $t = 0$.  In
addition equation \eref{epsilon_scaling} is the lowest-order
contribution in $\sigma \sqrt{\frac{t}{N}}$ from an asymptotic series
generated from an iterative solution to \eref{GBM} (manuscript in
preparation).

The approximation in \eref{lin_approx} neglects any non-linear
effects. Nonetheless it is informative of the trade-off between $N$
and $t$.  We can set \eref{epsilon_scaling} equal to some finite value
and derive an expression for the time, $\tau$, it takes to reach
this deviation for a given ensemble size $N$. From
\eref{epsilon_scaling}, this is
\begin{equation}
\tau \sim \frac{1}{\mu} \left(\ln\left(\epsilon_N(t)\sqrt{N}\right)-\ln\left(\sigma\sqrt{\tau}\right)\right)
\elabel{scaling}
\end{equation}
Compared to the logarithmic scaling, $\tau \sim \ln(N)$,
\eref{scaling} includes a correction (the second term on the
right-hand side). For large characteristic times, $\tau \gg
|\ln\left(\sigma\sqrt{\tau}\right)|$ (\ie large values of
$\epsilon_N(t)\sqrt{N})$, we can neglect this correction. Note
however, that for the asymptotic expansion to be a valid approximation
we must have $\sigma \sqrt{t/N}$ is small for $t=\tau$.

In \fref{abs_dev} we show the times $\tau$ where absolute
deviations of magnitude 0.1 and 1 of the PEA from the ensemble average
were reached in the trajectories in \fref{trajectories}.

Considering the asymptotic series from the iterative solution to
\eref{GBM}, we note that the amplitude of the second-order term
divided by the amplitude of the first-order term for the data in
\fref{abs_dev} is less than $0.55$ for $N\geq 10$.  For $\epsilon_N(t)
= 1$, where the $\ln(N)$-scaling appears to be valid, this ratio is
small, namely ranging from 0.55 for $N = 10$ to 0.08 for $N=1000$.  It
is even smaller when $\epsilon_N(t)=0.1$. This implies that the linear
approximation is a good description for a wide range of parameters.

\begin{figure}[h]
\includegraphics*[width=\columnwidth,angle=0]{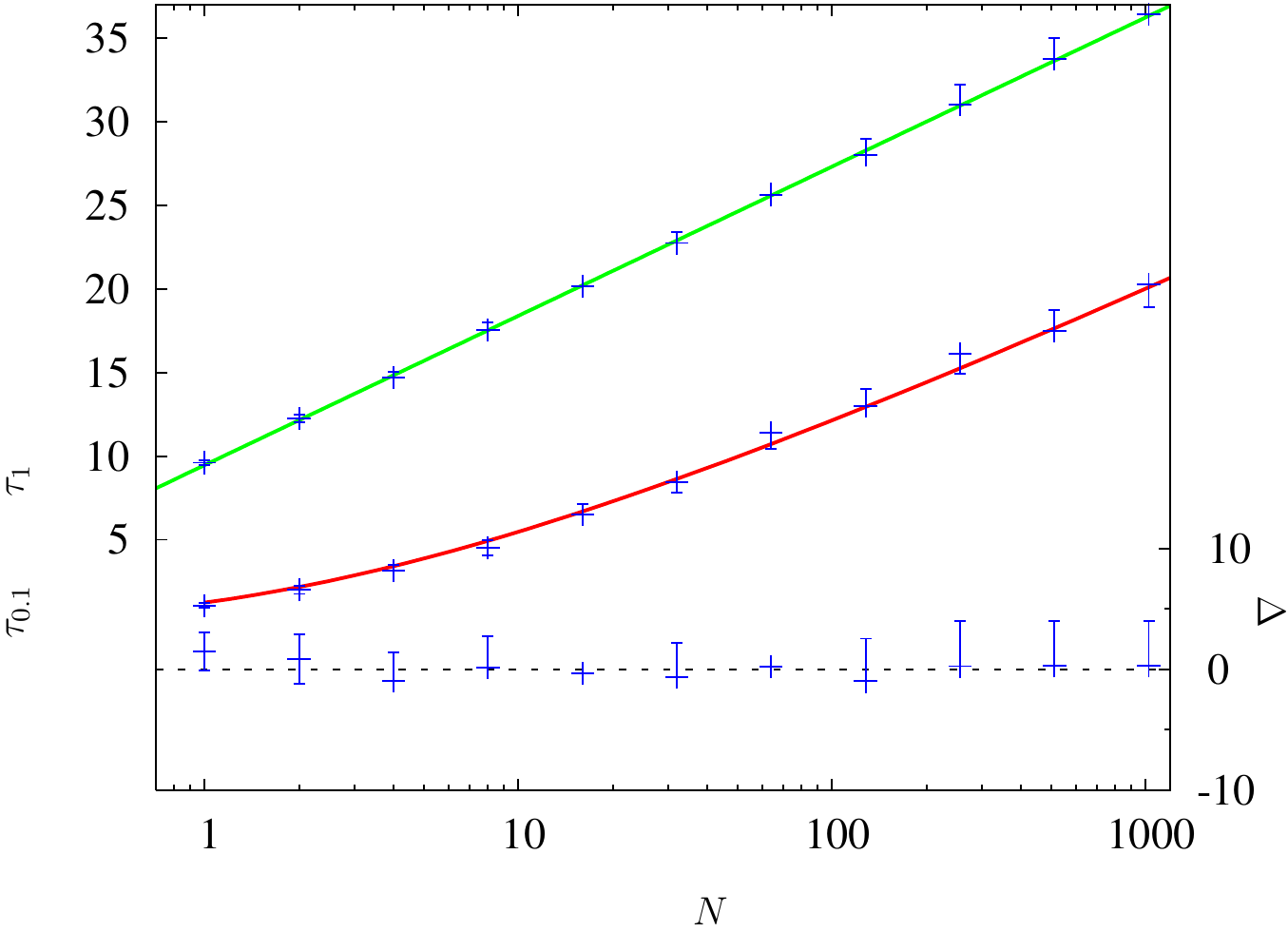}
\caption{Time where the median for a given $N$, shown in
  \fref{trajectories}, first differs from the ensemble average
  $\exp(\mu t)$ by at least 1.0 ($\tau_{1}$, top) and 0.1
  ($\tau_{0.1}$, middle). Error bars show the corresponding times for
  the 49th and 51st percentile. A purely logarithmic fit of the form
  $\tau=a+b\ln(N)$ works well for deviation 1.0 (with $a=9.46$ and
  $b=3.88$). For deviation 0.1 occurring at smaller $t$ corrections
  become visible; to compare shapes the solution for $\tau(N)$ of
  \eref{scaling} is shown in arbitrary units. Bottom: the percentage
  difference, $\Delta$, between the top data and their fit.}
\flabel{abs_dev}
\end{figure}

Also features associated with much larger deviations, such as
zero-crossings of the growth rate, \eref{estimator}, approximately
follow logarithmic scaling. This can be seen in \fref{trajectories},
where the spacing of successive zero-crossings of the growth rate (\ie
trajectories crossing 1) is approximately constant for each doubling
of $N$. For a small deviation $\epsilon_N(t)=0.1$ we find a
non-logarithmic shape similar to \eref{scaling} including the
correction.

The fact that properties of a linear approximation feed through to the
full non-linear solution is surprising but not
unheard-of. \Eref{escale-eq} is identical to the Cahn-Hilliard-Cook
theory for systems with a non-conserved order parameter which
describes the evolution of a class of materials after a quench into an
unstable state \cite{GrossKleinLudwig1994}.  The Cahn-Hilliard-Cook
theory, like \eref{escale-eq}, is an early-time linear theory that
accurately describes the sensitivity of the system to arbitrary
perturbations.  The implication is that early growth associated with
PEAs of GBMs is inherently unstable. This leads to the conclusion that
any PEA will eventually be dominated by the same time-average behavior
(\eref{time} holds for arbitrary $N$).

In economics a mistaken belief in ergodicity has produced wide-spread
conceptual inconsistency. For instance while ergodic models of
exchange yield realistic predicitons for the lower part of wealth
distributions, it has been pointed out that GBM-like multiplicative
non-ergodic models are most natural for the upper part
\cite{YakovenkoRosser2009}. Under GBM, the so-called Theil index of
inequality \cite{Theil1967} can be viewed as the time-integrated
difference between the time-average and ensemble-average growth
rates. This difference (i.e. inequality) would be zero if GBM were
ergodic. As more sophisticated and realistic economic models are
studied \cite{YakovenkoRosser2009} it will be important to understand
the relation between the nature of the noise, the presence of
ergodicity and the properties of PEAs. Our results have important
implications for the relevance of diversification strategies under
realistic conditions, and the effect of multiplicative noise, in
fields ranging from financial risk management to ecology, evolutionary
biology, and material science.

\acknowledgments{This work was supported by EPSRC Mathematics Platform
  grant EP/I019111/1, ZONlab ltd., and the DOE through grant
  DE-FG02-95ER14498.}

\bibliography{/Users/obp48/bibliography/bibliography}

\begin{thebibliography}{13}
\expandafter\ifx\csname natexlab\endcsname\relax\def\natexlab#1{#1}\fi
\expandafter\ifx\csname bibnamefont\endcsname\relax
  \def\bibnamefont#1{#1}\fi
\expandafter\ifx\csname bibfnamefont\endcsname\relax
  \def\bibfnamefont#1{#1}\fi
\expandafter\ifx\csname citenamefont\endcsname\relax
  \def\citenamefont#1{#1}\fi
\expandafter\ifx\csname url\endcsname\relax
  \def\url#1{\texttt{#1}}\fi
\expandafter\ifx\csname urlprefix\endcsname\relax\def\urlprefix{URL }\fi
\providecommand{\bibinfo}[2]{#2}
\providecommand{\eprint}[2][]{\url{#2}}

\bibitem[{\citenamefont{Peters}(2011{\natexlab{a}})}]{Peters2011a}
\bibinfo{author}{\bibfnamefont{O.}~\bibnamefont{Peters}},
  \bibinfo{journal}{Quant. Fin.} \textbf{\bibinfo{volume}{11}},
  \bibinfo{pages}{1593} (\bibinfo{year}{2011}{\natexlab{a}}),
  \href{http://dx.doi.org/10.1080/14697688.2010.513338}{doi:10.1080/14697688.2010.513338}.

\bibitem[{\citenamefont{Chernoff and Moses}(1959)}]{ChernoffMoses1959}
\bibinfo{author}{\bibfnamefont{H.}~\bibnamefont{Chernoff}} \bibnamefont{and}
  \bibinfo{author}{\bibfnamefont{L.~E.} \bibnamefont{Moses}},
  \emph{\bibinfo{title}{Elementary Decision Theory}} (\bibinfo{publisher}{John
  Wiley \& Sons}, \bibinfo{year}{1959}).

\bibitem[{\citenamefont{Lau and Lubensky}(2007)}]{LauLubensky2007}
\bibinfo{author}{\bibfnamefont{A.~W.~C.} \bibnamefont{Lau}} \bibnamefont{and}
  \bibinfo{author}{\bibfnamefont{T.~C.} \bibnamefont{Lubensky}},
  \bibinfo{journal}{Phys. Rev. E} \textbf{\bibinfo{volume}{76}},
  \bibinfo{pages}{011123} (\bibinfo{year}{2007})
  \href{http://dx.doi.org/10.1103/PhysRevE.76.011123}{doi:10.1103/PhysRevE.76.011123}.

\bibitem[{\citenamefont{Gray}(2009)}]{Gray2009}
\bibinfo{author}{\bibfnamefont{R.~M.} \bibnamefont{Gray}},
  \emph{\bibinfo{title}{Probability, random processes, and ergodic properties}}
  (\bibinfo{publisher}{Springer}, \bibinfo{year}{2009}), \bibinfo{edition}{2nd}
  ed.

\bibitem[{\citenamefont{{Kelly~Jr.}}(1956)}]{Kelly1956}
\bibinfo{author}{\bibfnamefont{J.~L.} \bibnamefont{{Kelly~Jr.}}},
  \bibinfo{journal}{Bell Sys. Tech. J.} \textbf{\bibinfo{volume}{35}},
  \bibinfo{pages}{917} (\bibinfo{year}{1956}).

\bibitem[{\citenamefont{Peters}(2011{\natexlab{b}})}]{Peters2011b}
\bibinfo{author}{\bibfnamefont{O.}~\bibnamefont{Peters}},
  \bibinfo{journal}{Phil. Trans. R. Soc. A} \textbf{\bibinfo{volume}{369}},
  \bibinfo{pages}{4913} (\bibinfo{year}{2011}{\natexlab{b}}),
  \href{http://dx.doi.org/10.1098/rsta.2011.0065}{doi:10.1098/rsta.2011.0065}.


\bibitem[{\citenamefont{Gumbel}(1958)}]{Gumbel1958}
\bibinfo{author}{\bibfnamefont{E.~J.} \bibnamefont{Gumbel}},
  \emph{\bibinfo{title}{Statistics of Extremes}} (\bibinfo{publisher}{Columbia
  University Press}, \bibinfo{year}{1958}).

\bibitem[{\citenamefont{Richardson}(1921)}]{Richardson1921}
\bibinfo{author}{\bibfnamefont{L.~F.} \bibnamefont{Richardson}},
  \bibinfo{journal}{Phil. Trans. Roy. Soc. A} \textbf{\bibinfo{volume}{221}},
  \bibinfo{pages}{1} (\bibinfo{year}{1921})
  \href{http://dx.doi.org/10.1098/rsta.1921.0001}{doi:10.1098/rsta.1921.0001}.


\bibitem[{\citenamefont{Redner}(1990)}]{Redner1990}
\bibinfo{author}{\bibfnamefont{S.}~\bibnamefont{Redner}}, \bibinfo{journal}{Am.
  J. Phys.} \textbf{\bibinfo{volume}{58}}, \bibinfo{pages}{267}
  (\bibinfo{year}{1990})
  \href{http://dx.doi.org/10.1119/1.16497}{doi:10.1119/1.16497}.

\bibitem[{\citenamefont{Morse and Feshbach}(1953)}]{MorseFeshbach1953}
\bibinfo{author}{\bibfnamefont{P.~M.} \bibnamefont{Morse}} \bibnamefont{and}
  \bibinfo{author}{\bibfnamefont{H.}~\bibnamefont{Feshbach}},
  \emph{\bibinfo{title}{Methods of Theoretical Physics}}
  (\bibinfo{publisher}{McGraw-Hill}, \bibinfo{year}{1953}).

\bibitem[{\citenamefont{Gross et~al.}(1994)\citenamefont{Gross, Klein, and
  Ludwig}}]{GrossKleinLudwig1994}
\bibinfo{author}{\bibfnamefont{N.}~\bibnamefont{Gross}},
  \bibinfo{author}{\bibfnamefont{W.}~\bibnamefont{Klein}}, \bibnamefont{and}
  \bibinfo{author}{\bibfnamefont{K.}~\bibnamefont{Ludwig}},
  \bibinfo{journal}{Phys. Rev. Lett.} \textbf{\bibinfo{volume}{73}},
  \bibinfo{pages}{2639} (\bibinfo{year}{1994})
  \href{http://dx.doi.org/10.1103/PhysRevLett.73.2639}{doi:10.1103/PhysRevLett.73.2639}.

\bibitem[{\citenamefont{Yakovenko and {Rosser,
  Jr.}}(2009)}]{YakovenkoRosser2009}
\bibinfo{author}{\bibfnamefont{V.~M.} \bibnamefont{Yakovenko}}
  \bibnamefont{and} \bibinfo{author}{\bibfnamefont{J.~B.} \bibnamefont{{Rosser,
  Jr.}}}, \bibinfo{journal}{Rev. Mod. Phys.} \textbf{\bibinfo{volume}{81}},
  \bibinfo{pages}{1703} (\bibinfo{year}{2009})
  \href{http://dx.doi.org/10.1103/RevModPhys.81.1703}{doi:10.1103/RevModPhys.81.1703}.


\bibitem[{\citenamefont{Theil}(1967)}]{Theil1967}
\bibinfo{author}{\bibfnamefont{H.}~\bibnamefont{Theil}},
  \emph{\bibinfo{title}{Economics and information theory}}
  (\bibinfo{publisher}{North-Holland Publishing Company},
  \bibinfo{year}{1967}).

\end{thebibliography}
\end{document}